\documentclass[preprint2]{aastex}
\usepackage{emulateapj5}
\usepackage{onecolfloat5}

\def\Sref#1{\S\ref{Sec:#1}}
\def\Fref#1{Figure~\ref{Fig:#1}}
\def\Tref#1{Table~\ref{Table:#1}}
\newcommand{\vperp} {V_{\perp}}
\newcommand{\pmra}  {\mu_{\alpha}}
\newcommand{\pmdec} {\mu_{\delta}}
\newcommand{\persec}{s$^{-1}$}
\newcommand{\peryr} {yr$^{-1}$}
\newcommand{\maspyr}{mas~yr$^{-1}$}
\newcommand{\qinb}  {J1934+1043} 
\newcommand{\dof}   {degrees of freedom}
\newcommand{\zetaoph}{$\zeta$~Oph}
\newcommand{\aam}   {\altaffilmark}

\long\def\symbfoottxt[#1]#2{\begingroup%
\def\thefootnote{\fnsymbol{footnote}}\footnotetext[#1]{#2}\endgroup} 

\slugcomment{Version 3.0, 23 Nov 2003}
\shorttitle{Pulsar Parallaxes at 5 GHz}
\shortauthors{Chatterjee et al.}

\begin{document}
\twocolumn[
\title{Pulsar Parallaxes at 5 GHz with the Very Long Baseline Array}
\author{S. Chatterjee\aam{1,2,\dag}, J. M. Cordes\aam{1}, 
        W. H. T. Vlemmings\aam{1}, Z. Arzoumanian\aam{3}, \\
        W. M. Goss\aam{2}, and T. J. W. Lazio\aam{4}}

\begin{abstract}
We present the first pulsar parallaxes measured with phase-referenced
pulsar VLBI observations at 5~GHz. Due to the steep spectra of
pulsars, previous astrometric measurements have been at lower frequencies.
However, the strongest pulsars can be observed at 5~GHz, offering the
benefit of lower combined ionospheric and tropospheric phase errors,
which usually limit VLBI astrometric accuracy. The pulsars B0329+54,
B0355+54 and B1929+10 were observed for 7 epochs spread evenly over 2
years. For B0329+54, large systematic errors lead to only an upper
limit on the parallax ($\pi < 1.5$~mas).  A new proper motion and
parallax were measured for B0355+54 ($\pi = 0.91 \pm 0.16$~mas),
implying a distance of $1.04^{+0.21}_{-0.16}$~kpc and a transverse
velocity of $61^{+12}_{-9}$~km~\persec.  The parallax and proper
motion for B1929+10 were significantly improved ($\pi = 2.77 \pm
0.07$~mas), yielding a distance of $361^{+10}_{-8}$~pc and a
transverse velocity of $177^{+4}_{-5}$~km~\persec.  We demonstrate
that the astrometric errors are correlated with the angular separation
between the phase reference calibrator and the target source, with
significantly lower errors at 5~GHz compared to 1.6~GHz.  Finally,
based on our new distance determinations for B1929+10 and B0355+54, we
derive or constrain the luminosities of each pulsar at high energies.
We show that, for thermal emission models, the emitting area for
X-rays from PSR B1929+10 is roughly consistent with the canonical size
for a heated polar cap, and that the conversion of spin-down power to
gamma-ray luminosity in B0355+54 must be low.  The new proper motion
for B1929+10 also implies that its progenitor is unlikely to have been
the binary companion of the runaway O-star $\zeta$~Ophiuchi.

\end{abstract}
\keywords{astrometry --- pulsars:individual (PSR B0355+54, 
  PSR B1929+10) --- stars:kinematics --- stars:neutron --- X-rays:stars} 
]

\altaffiltext{1}{Department of Astronomy, Cornell University, Ithaca, NY 14853; 
                 shami@astro.cornell.edu}
\altaffiltext{2}{National Radio Astronomy Observatory, P.O. Box O, Socorro, NM 87801.}
\symbfoottxt[2]{Jansky Postdoctoral Fellow}
\altaffiltext{3}{USRA, Laboratory for High-Energy Astrophysics, NASA-GSFC, Code 662,
Greenbelt, MD 20771.}
\altaffiltext{4}{Naval Research Laboratory, Code 7213, Washington, DC 20375.}

\section{Introduction}\label{Sec:intro}

The measurement of distances is a fundamental problem in astronomy.
Trigonometric parallax offers one of the only model-independent
mechanisms for such measurements, thus defining the lowest rungs of
the distance ladder. The parallaxes and proper motions of pulsars,
obtained through imaging or pulse-timing observations, provide
measurements of their distances and velocities, and thus allow us to
probe the physics of neutron star formation in supernovae, as well as
the intervening interstellar medium which scatters and disperses the
radio pulses from pulsars.

Pulsar distances provide crucial calibration for electron density
distribution models of the Galaxy \citep[e.g.][]{TC93,CL02}, which, in
turn, provide estimates of the distances for the majority of the
pulsar population and underlie estimates of their velocities.  The
velocity distribution of pulsars \citep{ACC02} helps constrain core
collapse processes in supernovae \citep{BH96}, while velocities of
individual objects can be used to trace their progenitor supernovae
and birth sites \citep{HBZ01}.
Recent results from X-ray observations, coupled with distance
estimates, have established constraints on neutron star cooling curves
\citep{TTT+02} which limit plausible nuclear Equations of State. 

While the diverse applications of pulsar distances and velocities make
a compelling case for the necessity of pulsar astrometry, reliable
measurements have been few and far between \citep[see,
e.g.][]{TBM+99}.  The paucity of parallax measurements is due to the
difficulty of measuring the effect. For a pulsar at 1~kpc, the angular
displacement is only 1~mas and the semi-annual timing perturbation is
1.6~$\mu$s. Few pulsars allow sufficient precision to measure the
timing effect, while astrometry with {\em ad hoc} very long baseline
interferometry (VLBI) arrays has been difficult and inconsistent.
However, the advent of the NRAO Very Long Baseline Array, a full-time,
dedicated VLBI instrument, coupled with pulsar gating and refined
calibration techniques, has significantly altered the situation and
led to several parallax measurements in recent years
\citep{BBB+00,CCL+01,BBGT02}. Due to the steep spectra of pulsars
($S_{\nu} \propto \nu^{\alpha}, \langle\alpha\rangle \sim -1.6$;
\citealt{LYLG95}) these measurements have all been at 1.4~GHz, a
frequency lower than that used for most other VLBI astrometry. Here we
present the first pulsar parallaxes measured at 5~GHz. In \Sref{obs},
we outline the observations, calibration process and data
analysis. Astrometric results for the pulsars B0329+54, B0355+54 and
B1929+10 are presented in \Sref{results}, and compared to those of
\citet{BBGT02} for B0329+54 and B1929+10.  We estimate the attainable
astrometric accuracy at 1.4~GHz and 5~GHz in \Sref{cal}, and discuss the
implications of the measured distances and velocities for B1929+10 and
B0355+54 in \Sref{imply}.

\section{Observations and Data Analysis}\label{Sec:obs}

Astrometric VLBI observations require the visibility phase of the
target source to be calibrated using observatons of a nearby
calibrator source.  Typically, the target and the ``nodding''
calibrator are observed in short alternate scans. To achieve phase
connection across multiple scans, phase referencing requires rapid
cycle times and a small angular separation between the target and the
nodding calibrator (hereafter referred to as the ``calibrator
throw'').  The limiting source of astrometric error is typically the
uncalibrated differential atmosphere between the target and the
calibrator, which includes an ionospheric dispersive phase error
($\propto \nu^{-1}$) and a tropospheric non-dispersive phase error
($\propto \nu$).  These two error terms are comparable in size at the
5~GHz band and provide an overall approximate minimum error, making
the band well suited for astrometry. For example, a 0.36~mas parallax
has been measured for Sco~X-1 by \citet{BFG99} at 5~GHz. We chose
three pulsars, B0329+54, B0355+54 and B1929+10, which have flux
densities between 1 and 10~mJy at 5~GHz, and observed them for 7
epochs separated by 4 months each between 2000 March and 2002 March.
Solar maximum occurred within this period, leading to heightened
ionospheric activity and $\sim$ 3--10 $\times$ increases in the daily
average total electron content compared to that at solar minimum as
measured by GPS \citep[e.g.][]{solarmax}.  Such ionospheric activity
is expected to cause epoch-dependent systematic errors in our
astrometry. 

At each epoch, the pulsars were phase-referenced to extragalactic
sources included in the International Celestial Reference Frame.
B0329+54 and B0355+54 were phase-referenced to ICRF~J035929.7+505750
(NRAO~150), from which they are separated by 5.3\arcdeg\ and
3.3\arcdeg, respectively.  For B1929+10, ICRF~J192840.8+084848, a
calibrator 2.3\arcdeg\ away, was used as a reference source, and each
observation included a few scans on ICRF~J193435.0+104340 (hereafter
\qinb), only 0.6\arcdeg\ from the pulsar. The absolute positions of
the calibrator sources are known to between 2--3~mas \citep{MAE+98}.
The observations alternated between 90 seconds on the
calibrator and 150 seconds on the target, and the pulsar gate was
applied to the data at the VLBA correlator. Gating boosted the S/N
ratio by $f^{-1/2}$, a factor of $\sim$ 3--4, where $f = T_{\rm on} /
(T_{\rm on} + T_{\rm off})$ is the gate duty cycle. AIPS, the
Astronomical Image Processing System, was used for the data reduction,
which followed standard procedures \citep{conf-BC95}. The basic steps
include flagging of bad records, amplitude calibration based on the
system temperatures at each antenna, fringe fitting, and several
passes of self calibration on the nodding calibrator.  The multiple
self calibration passes serve to correct the visibility phases and
gains while fixing the position of the nodding calibrator.

For B0329+54 and B0355+54, the calibration solutions derived for the
nodding calibrator were transferred to the pulsar data, images were
created, and the AIPS task {\tt JMFIT} was used to obtain astrometric
positions at each epoch. As described by \citet{CCL+01}, position
uncertainties were derived as a combination of the synthesized beam
resolution and systematic errors in quadrature, where the systematic
errors were estimated at each epoch using the deviation of the
deconvolved image from that expected for a point source.  While the
typical uncertainties are $<1$~mas, the systematic errors were worse
for B0329+54 at every epoch, and exceeded 5 mas at one epoch (2000
November, due to ionospheric activity) for both pulsars, rendering
that epoch useless for astrometry.

For B1929+10, after transferring the calibration solutions from the
nodding calibrator, images were created for both the pulsar and the
second reference source \qinb, and {\tt JMFIT} was used as
above. Ideally, a constant position should be obtained for \qinb\ at
each epoch, but the observed position varied randomly, with a median
angular scatter (i.e., the median of the angular position offsets)
of 0.7 mas over 7 epochs. Phase self-calibration of
\qinb\ fixes its position at the expected location, and transferring
these solutions to B1929+10 not only shifts the observed pulsar
position, but also improves the image quality and reduces the
estimated systematic errors significantly. Thus the astrometric
positions for B1929+10 were referenced to \qinb, 0.6\arcdeg\
away. Here, too, position uncertainties at one epoch (2001 March)
exceeded 5 mas, and we discarded it in the final astrometric fits.

\section{Parallaxes and Proper Motions}\label{Sec:results}

The derived positions and position uncertainties for each pulsar at
5~GHz were used to fit for an annual trigonometric parallax, along
with zero-point position offsets and proper motions in each coordinate
(5 parameters), using a weighted least-squares analysis. For B0329+54
and B1929+10, positions measured by \citet{BBGT02} at 1.4~GHz were
also used to perform joint fits: two independent zero-point position
offsets were employed for the two data sets to account for the
different nodding calibrators and observing frequencies employed. We
describe the results for each pulsar in turn.

For B0329+54, unmodeled systematic errors dominate the fit and no
satisfactory solution was obtained. The formal value of the parallax
is $\pi = -0.45 \pm 0.36$ mas, with a reduced $\chi^2 = 2.1$ for 7
\dof. Holding the parallax fixed at 0 does not improve the fit for
proper motion, yielding $\pmra = 16.8 \pm 0.4$~\maspyr, $\pmdec =
-11.1 \pm 0.9$~\maspyr, and a reduced $\chi^2 = 2.3$ (8 \dof). Given
the position uncertainties, we estimate an upper limit on the parallax
$\pi < 1.5$ mas, which is consistent with the \citet{BBGT02} value
of $ 0.94 \pm 0.11$ mas. In a joint fit to both data sets, the larger
position uncertainties for the 5~GHz data leave the earlier parallax
unimproved. The joint proper motion is also within the errors of the
earlier Brisken et al.\ value of $\pmra = 17.00 \pm 0.27$~\maspyr\ and
$\pmdec = -9.48 \pm 0.37$~\maspyr. The 5~GHz data do not contribute
useful astrometry in this case, because the calibrator was too far
away (5.3\arcdeg) to adequately correct phase perturbations in the
data, as we discuss below.

The position uncertainties are better behaved for B0355+54 and a good
weighted least squares fit was obtained for the parallax and proper
motion, with a reduced $\chi^2$ = 0.82 for 7 \dof. The derived
best-fit values are listed in \Tref{results}, and imply most probable
values of $1.04^{+0.21}_{-0.16}$~kpc and $61^{+12}_{-9}$~km~\persec\
for the distance and the velocity, as listed in \Tref{derived}.
\Fref{0355} shows the residual position offsets in each coordinate
after subtracting the best fit proper motion, with sinusoids
corresponding to the best fit parallax overplotted. It is apparent
from the figure that the residual scatter in declination is
significantly worse than that in right ascension. The angular
separation between the calibrator and the pulsar is primarily in
declination ($\Delta\alpha_{\rm cal} = 0.1\arcdeg$, $\Delta\delta_{\rm
cal} = 3.3\arcdeg$), which may explain the excess scatter. Our
estimates of position uncertainty do not currently account for the
asymmetric calibrator throw, an issue we return to in \Sref{cal}.  In
tests of the fit after dropping one epoch at a time, the obtained
parallaxes range from $\pi = 0.87$~mas to $1.02$~mas, attesting to the
robustness of the reported value of $0.91 \pm 0.16$~mas for the
parallax.

For B1929+10, too, a good fit was obtained with a reduced $\chi^2 =
1.06$ for 7 \dof: the best-fit values are listed in \Tref{results}.
The longer time baseline of the 5~GHz data (2 years compared to 0.77
year for the 1.4~GHz data) results in a more accurate proper motion
determination, and yields a parallax $\pi = 2.76 \pm 0.14$~mas, which
differs from the \citet{BBGT02} value of $3.02 \pm 0.09$~mas at the
$2\sigma$ level.  The covariances between ($\pmra,\pmdec$) and $\pi$
are ($-0.37, -0.01$) for the 5~GHz fit, much lower values compared to
($0.79, 0.32$) for the 1.4~GHz data.  A joint fit to both data sets
yields a parallax that is essentially unchanged from the 5~GHz value,
but with smaller errors: $\pi = 2.77 \pm 0.07$~mas, with a reduced
$\chi^2 = 0.97$ for 15 \dof\ and covariances ($0.15, 0.14$) between
($\pmra,\pmdec$) and $\pi$.  The zero-point offsets for the two data
sets are expected to differ since they are referenced to different
nodding calibrators at different frequencies (J1945+0952 at 1.4~GHz
and ICRF~J192840.8+084848 at 5~GHz), but we obtain exceptionally good
agreement. For the best joint fit, the difference in the zero-point
offsets ($\Delta\alpha = 0.6$~mas, $\Delta\delta = 2.4$~mas) implies
that the {\em absolute} positions agree to $\sim 2.5$~mas. The
residual position offsets after subtracting the best fit proper motion
are shown in \Fref{1929}, with the best fit parallax overplotted. The
most probable value of the distance is $0.361^{+0.010}_{-0.008}$~pc,
and the velocity is $177^{+4}_{-5}$~km~\persec.
\Tref{derived} summarizes these and other derived parameters, and we
defer a discussion of the ramifications of these measured parameters
to \Sref{imply}.  The joint fits reported here are robust compared to
previous measurements reported for the same pulsar (\Tref{1929}), and
the gradual convergence of the measurements is also gratifying.


\begin{deluxetable}{lccccc}
\tablecolumns{6}
\tablewidth{0pc} 
\tablecaption{Measured Positions, Proper Motions and Parallaxes\label{Table:results}}
\tablehead{ 
\colhead{Pulsar} & \colhead{$\alpha_{2000}$\tablenotemark{\dag}} &
                   \colhead{$\delta_{2000}$\tablenotemark{\dag}} &
\colhead{$\pmra$} & \colhead{$\pmdec$} & \colhead{$\pi$} \\
\colhead{} & \colhead{} & \colhead{} &
\colhead{(mas \peryr)} & \colhead{(mas \peryr)} & \colhead{(mas)} 
}
\startdata
B0355+54 &
 $ 03^{\rm h}58^{\rm m}53\fs71650 $ & $ +54\arcdeg13\arcmin13\farcs7273 $ &
 $ 9.20 \pm 0.18 $ & $ 8.17 \pm 0.39 $ & $ 0.91 \pm 0.16 $ \\ 
B1929+10 &
 $ 19^{\rm h}32^{\rm m}13\fs94970 $ & $ +10\arcdeg59\arcmin32\farcs4198 $ &
 $ 94.03 \pm 0.14 $ & $ 43.37 \pm 0.29 $ & $ 2.76 \pm 0.14 $ \\ 
B1929+10 joint &
 $ 19^{\rm h}32^{\rm m}13\fs94969 $ & $ +10\arcdeg59\arcmin32\farcs4203 $ &
 $ 94.09 \pm 0.11 $ & $ 42.99 \pm 0.16 $ & $ 2.77 \pm 0.07 $ \\ 
\enddata 
\tablenotetext{\dag}{Absolute positions extrapolated to epoch 2000.0
 are accurate to $\sim 5$ mas, which includes the uncertainties in
 the fit position and extrapolation based on proper motion ($\sim
 1$~mas), as well as absolute position uncertainties ($<3$~mas) for the
 ICRF sources used in phase referencing.} 
\end{deluxetable} 


\begin{figure}[htf]
\plotone{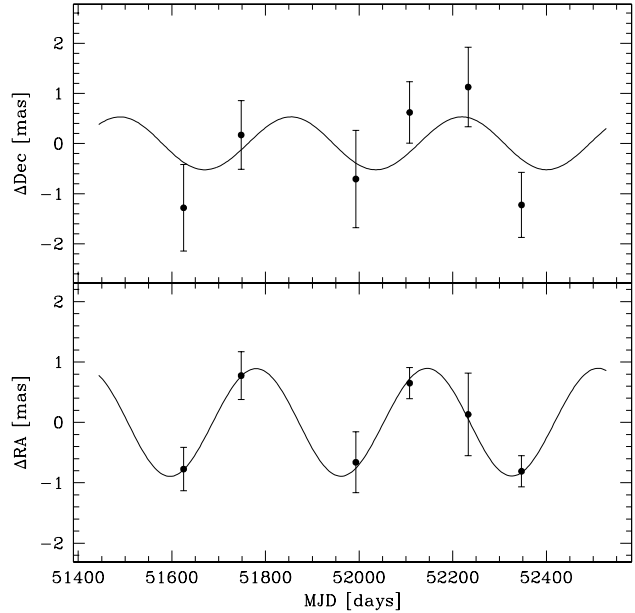}
\caption{The parallax signature of PSR~B0355+54 in right ascension and
  declination, after subtracting the best-fit proper motion from the
  astrometric positions. Sinusoids corresponding to the best fit
  parallax $\pi = 0.91$~mas are overplotted.}
\label{Fig:0355}
\end{figure}

\begin{figure}[htf]
\plotone{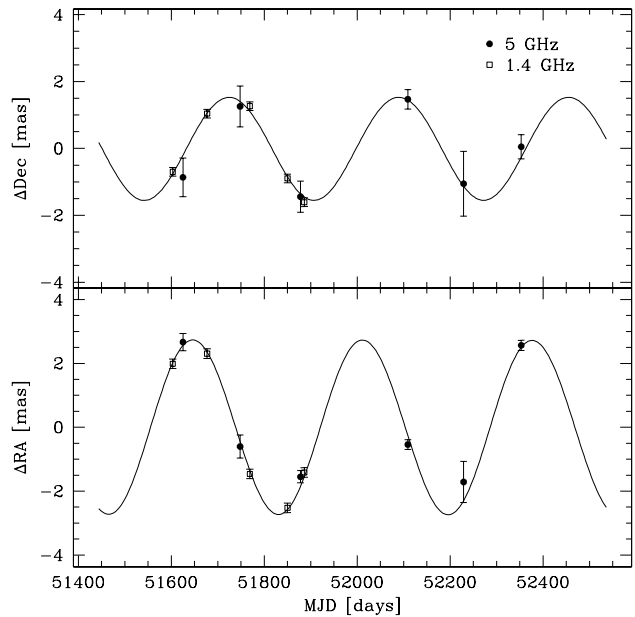}
\caption{The parallax signature of PSR~B1929+10 in right ascension and
  declination, after subtracting the best-fit proper motion from
  the astrometric positions measured at 5~GHz (closed circles; this
  work) and 1.4~GHz \citep[open squares;][]{BBGT02}. Sinusoids
  corresponding to the best fit parallax $\pi = 2.77$~mas are
  overplotted.}
\label{Fig:1929}
\end{figure}

\begin{deluxetable}{cccccc}
\tablecolumns{6}
\tablewidth{0pc} 
\tablecaption{Derived Parameters\label{Table:derived}}
\tablehead{ 
\colhead{Pulsar} & \colhead{DM} & \colhead{Distance} & 
\colhead{D$_{\rm NE2001}$} & \colhead{V$_{\perp}$} & 
\colhead{$\langle n_e \rangle$} \\
\colhead{} & \colhead{(pc cm$^{-3}$)} & \colhead{(kpc)} & 
\colhead{(kpc)} & \colhead{(km \persec)} & 
\colhead{(cm$^{-3}$)}
}
\startdata
B0355+54 & 57.14 & $1.04^{+0.21}_{-0.16}$  & 1.45 & $61^{+12}_{-9}$  & 0.052  \\
B1929+10 & 3.176 & $0.361^{+0.010}_{-0.008}$ & 0.33 & $177^{+4}_{-5}$ & 0.0087 \\
\enddata 
\end{deluxetable}

\begin{deluxetable}{lcl}
\tablecolumns{3}
\tablewidth{0pc} 
\tablecaption{Measured Parallaxes for B1929+10\label{Table:1929}}
\tablehead{ 
\colhead{Band}  & \colhead{$\pi$} & \colhead{Reference} \\
\colhead{(GHz)} & \colhead{(mas)} & \colhead{} 
}
\startdata
0.4     & $ 21.5 \pm 0.3$ & 1 \\
2.7     & $ <4 $          & 2 \\
2.2     & $ 5.0 \pm 1.5 $ & 3 \\
1.4     & $ 3.02 \pm 0.09 $ & 4 \\
5       & $ 2.76 \pm 0.14 $ & This work  \\
1.4, 5  & $ 2.77 \pm 0.07 $ & Joint fit  \\
\enddata 
\tablerefs{(1) \citet{SLA79}; (2) \citet{BS82}; (3) \citet{C95thesis}; 
           (4) \citet{BBGT02} }
\end{deluxetable} 


\section{Estimating Astrometric Accuracy}\label{Sec:cal}

In phase-referenced observations, the phases of the target
visibilities are inferred from those of the nodding calibrator.  For
each antenna, fluctuations in the ionosphere and troposphere are
sampled on an angular scale equal to the separation between the target
and calibrator sources.  The level of fluctuation is characterized by
the phase structure function of the atmosphere \citep[see, for
example,][13.1 and 13.3]{TMS}.  For VLB arrays, the atmosphere is
quite different at each antenna, and the uncalibrated differential
phase for each visibility, which is a baseline-dependent quantity,
cannot be interpreted as a simple excess in ionospheric path length or
water vapor column density.  It is reasonable, however, to expect that
the effectiveness of phase referencing in calibrating target
visibilities (and thus the quality of target images) depends on the
angular separation between the target and the calibrator over which
phase solutions have to be extrapolated.  The temporal separation
between scans on the calibrator and target introduces the
time-variability of the atmosphere as another source of error, but
this is modest in comparison, especially for short cycle times when
phase connection is achieved across multiple scans.

For the 5~GHz observations described here, B0329+54 and B0355+54
shared the same nodding calibrator, and scans on the two pulsars were
interleaved in time. But as described in \Sref{results}, the quality
of astrometry differed markedly for the two, leading to no useful
result for one and a good sub-mas parallax for the other. The primary
source of error is the 5.3\arcdeg\ calibrator throw for B0329+54, as
compared to only 3.3\arcdeg\ for B0355+54.  The observed position for
the calibrator source \qinb\ also varied when phase referenced to the
nodding calibrator 2.4\arcdeg\ away, while the positions for B1929+10
were well modeled by a proper motion and parallax after referencing to
\qinb, 0.6\arcdeg\ away. We quantify the random astrometric error for
each pulsar at each epoch by subtracting the best fit proper motion
and parallax model and combining the residuals for each coordinate in
quadrature. The median error for each of the three target pulsars is
plotted against calibrator throw in \Fref{errors}, along with the
angular position scatter for \qinb\ about the median position. In each
case, the central 50\% range of the estimates is indicated with error
bars.  We note that this plot includes just 7 epochs, observed at only
two declinations and three sun angles, with potentially unusual
ionospheric activity at some epochs due to solar maximum. However, the
trend of increasing median errors with calibrator throw is immediately
apparent for the 5~GHz data.


\begin{figure}[htf]
\plotone{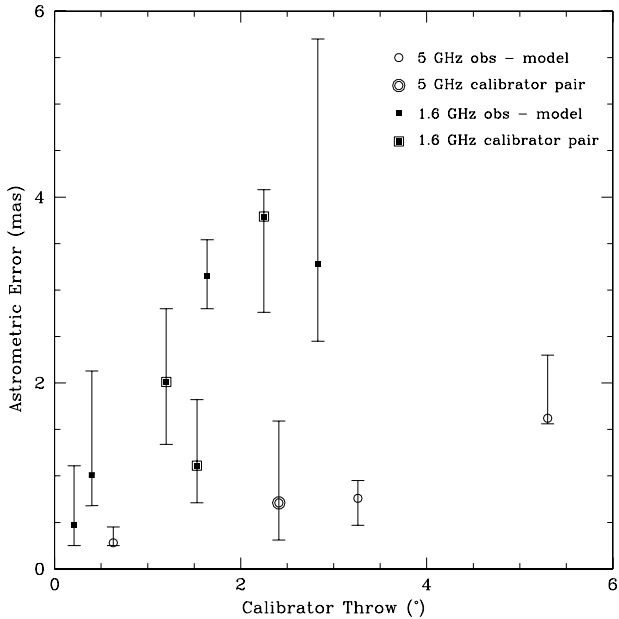}
\caption{Astrometric position errors as a function of the angular
  separation between the target and the calibrator source. The points
  indicate the median scatter in the position of one calibrator
  measured with respect to another (outlined), or the offset of an
  astrometric target position from that predicted by the best-fit
  proper motion and parallax model. The error bars indicate the inner
  50\% of the range in the scatter, between the first and third
  quartile. The 5~GHz data (open circles) are from this work. The
  1.6~GHz data (solid squares) include results from in-beam
  calibration for PSR~B0919+06 \citep[left-most point;][]{CCL+01} and
  VLBI astrometry of circumstellar OH masers \citep{VVD+03}. There are
  between 4 and 12 samples at each calibrator throw, limiting the
  statistical significance of this plot at present, but the increase
  in astrometric error with calibrator throw is an unmistakable
  trend. Astrometric accuracy in phase referenced observations is also
  better for 5~GHz data compared to 1.6~GHz data.}
\label{Fig:errors}
\end{figure}

For comparison, we also show astrometric data at 1.4--1.6~GHz from
prior projects. The parallax of PSR~B0919+06 has been measured using
in-beam calibration \citep{CCL+01}, effectively using a calibrator
throw of only 0.2\arcdeg.  The median residual random error, measured
as above, is $\sim 0.5$~mas.  The parallaxes of circumstellar masers
have been measured by \citet{VVD+03} using standard phase referencing
at 1.6~GHz, and the median residual errors range between 1--3~mas.
The median errors are indicated in \Fref{errors} with solid squares,
along with error bars showing the inner 50\% range. Additionally,
\citet{VVD+03} performed relative astrometry between calibrator pairs,
and we indicate the measured median relative scatter with outlined
squares. Caveats similar to those listed for the 5~GHz data also apply
here, since there are only between 4 and 12 epochs used for each
measurement. As in the case of the 5~GHz data, though, the astrometric
error at 1.6~GHz generally increases with increasing calibrator
throw. The errors are also significantly higher than those measured at
5~GHz for comparable calibrator throws.

\Fref{errors} empirically demonstrates the potential superiority of
ordinary phase-referenced astrometry at 5~GHz, and emphasizes the need
for additional techniques such as in-beam calibration
\citep{FGBC99,CCL+01} or wide-band ionospheric calibration
\citep{BBB+00} to obtain higher astrometric accuracy at lower
frequencies. In \Sref{results}, we noted that the position errors for
B0355+54 in each dimension ($\Delta\alpha,\Delta\delta$) appeared to
be correlated to the calibrator throw in that dimension
($\Delta\alpha_{\rm cal}, \Delta\delta_{\rm cal}$). This trend is
observed for all the objects in \Fref{errors}, at both 1.4 and
5~GHz. While the errors ($\Delta\alpha,\Delta\delta$) are unlikely to
be completely independent of each other, the calibrator throw in each
dimension is apparently an important predictor of the accuracy
attainable in that dimension.  Large programs now underway will
provide data on astrometric precision at currently unsampled
calibrator throws, as well as sampling a range of sun angles and a
larger span of the solar cycle, enhancing the utility of future
versions of this plot both as an observation planning tool and as a
probe of the tropospheric and ionospheric structure
functions. Meanwhile, the current results reiterate the importance of
using the smallest possible calibrator throw in phase-referenced
observations.  Additionally, based on these and other recent results
\citep{CCL+01,BBGT02}, we predict that parallaxes and proper motions
should be measurable for the majority of pulsars within 2~kpc, and
even for a some out to $\sim 3$~kpc, using either observations at
5~GHz, or lower frequency observations with additional calibration
techniques.  Beyond this distance range, the required precision may
run up against the limits imposed by potential calibrator source
structure and variability at the 100~$\mu$as level.

\section{Implications: B0355+54 and B1929+10}\label{Sec:imply}

The measured parallaxes and proper motions for B1929+10 and B0355+54
lead to estimates of the distance ($D$) and velocity perpendicular to
the line of sight ($\vperp$) for each pulsar.  The most probable
values for $D$ and $\vperp$ and the most compact 68\% probability
intervals are summarized in \Tref{derived}.  The distance estimated
from the dispersion measure (DM $= \int_0^D n_e(s)\,ds$) using the
NE2001 model \citep{CL02} is listed for comparison.  This model
already includes a distance constraint for B1929+10 based on the
Brisken et al.\ parallax and overestimates the measured distance to
B0355+54 by $\sim 40\%$.  \Tref{derived} also lists the average
electron density integrated over the line of sight, which is given
simply by ${\rm DM}/D$. The large difference between the two lines of
sight reflects the complex structure in the local interstellar medium,
as discussed, for example, by \citet{TBM+99} and \citet{CL02}.  

The structure of the local interstellar medium also affects the
scattering properties of the pulsar signal, and measured scattering
parameters can be used to constrain the distribution of scattering
material along the line of sight \citep[e.g.][]{CCL+01}. Preliminary
investigations suggest that the scattering properties of B1929+10 are
affected by interstellar material which may be located quite close to
the solar system. However, the velocity of the Earth is not negligible
compared to the measured velocities of B1929+10 and B0355+54, and we
postpone a detailed analysis of the interstellar scattering properties
of these pulsars to future work.

The astrometric results discussed here also lead us to re-examine, in
turn, the birth site for B1929+10 based on its kinematics, the nature
of its observed X-ray emission, and constraints on the cooling curve
for each of the two neutron stars.

\subsection{The Birth Site of B1929+10}

Based on numerical simulations, \citet{HBZ01} have suggested that
PSR~B1929+10 (which they refer to as J1932+1059) and the runaway
O-star $\zeta$~Ophiuchi were part of a binary system, which became
unbound when the pulsar progenitor exploded in a supernova $\sim
1$~Myr ago in the Upper Scorpius subgroup of the Sco OB2
association. They attribute the kick velocity of the pulsar and the
large space velocity of \zetaoph\ ($\sim 20$~km~\persec) to the same
supernova-induced binary disruption event.  As seen in Figures 4 and 5
of \citet{HBZ01}, only a limited region in the parameter space of
pulsar parallax, proper motion and radial velocity allows the pulsar
and \zetaoph\ to approach within 10~pc of each other and 10~pc of the
center of the Upper Scorpius association. For these successful
simulations (0.14\% of the total), the implied kinematic age of
B1929+10 is $\sim 1$~Myr, comparable to its spindown age of 3~Myr. 

The distance and velocity measurements underpinning such a birth
scenario are based on a ``dearth of data'' \citep{C95thesis}, and the
new values for the proper motion and parallax of B1929+10 that we
report here make the scenario extremely unlikely.  For $\pi =
2.76$~mas, a high radial velocity ($\gtrsim 300$~km~\persec) is
required, while our proper motion measurements are inconsistent with
the parameters used in the successful simulations by $>10\sigma$.
Therefore, B1929+10 is probably not the former binary companion of
\zetaoph. We note, however, that the origin of the pulsar in the Upper
Scorpius region is not excluded, and the pulsar could still have had
its birth in a supernova in this region 1--2~Myr ago \citep{D92}. The
isolated neutron star RX~J1856.5$-$3754 had initially been proposed as
a potential binary companion to \zetaoph\ in the Upper Scorpius
association \citep{W01}, but the revised distance to the neutron star
\citep{KVA02,WL02} rules this scenario out, since the closest approach
by RX~J1856.5$-$3754 to the association center occurred only $\sim
0.5$~Myr ago. Thus, for \zetaoph, the alternative birth scenario in
the Upper Centaurus Lupus subgroup 2--3~Myr ago \citep{VVD96} looks
more plausible.

\subsection{The Nature of X-ray Emission from B1929+10}

X-ray emission from PSR B1929+10 was detected in observations with the
{\em Einstein} \citep{conf-H83}, {\em ROSAT} \citep{YHH94}, and {\em
ASCA} \citep{WH97} observatories. The latter two datasets allow for
spectral analysis that can be used to constrain properties of the
X-ray emitting region. \citet{YHH94} fit the {\em ROSAT} data to a
model for blackbody emission, deriving a temperature of roughly
$3\times 10^6$~K (hereafter, 1 MK~$\equiv 10^6$~K). From {\em
ASCA}-SIS data, \citet{WH97} find comparably successful fits to a
single blackbody ($T = 5.14 \pm 0.53$~MK) and to a power-law spectrum.
Finally, \citet{2003astro.ph..2452W} perform spectral fits to combined
{\em ROSAT} and {\em ASCA} data (using in this case the {\em ASCA}-GIS
detector) spanning the energy range 0.1--10~keV; they find acceptable
fits to model spectra consisting of either a single power law or two
blackbody components with temperatures of 2 and $\sim 7$~MK. New
observations will be needed to ascertain whether the emission is
thermal or nonthermal in nature. If thermal, the existing data alone
would rule out an origin for the X-rays in ``cooling'' emission from
the entire surface of the neutron star, both because the implied
temperature is unexpectedly high for such an old pulsar
(characteristic age $\sim 3$~Myr) and because the emitting area would
be unphysically small.  Indeed, as discussed by \citet{WH97}, the
inferred emitting area
is nearly two orders of magnitude smaller than theoretical predictions
even for the size of a heated polar cap assuming a purely dipolar,
star-centered magnetic field.  \citet{WRHZ98} elaborate on this
finding, arguing that a small observed cap size may be evidence for a
low heating current from the magnetosphere, or an off-center dipolar
geometry, perhaps as a result of crustal plate motion on the neutron
star surface.

The implications of a new distance determination depend on whether the
emission from the neutron star is thermal or nonthermal in nature. If
the emission is assumed to be thermal, and accurately represented by
the single-blackbody result of \citet{WH97}, our new distance
measurement implies a bolometric luminosity\footnote[1]{Unless otherwise
noted, all luminosities, temperatures, and emitting areas are quoted
for a distant observer. Where adjustments are made for redshift at the
neutron star surface, canonical values $M = 1.4\,M_\sun$ and $R = 10$
km, are assumed.} $L = (2.74\pm0.37)\times 10^{30}$ ergs~s$^{-1}$, and
an emitting area $A = (6.8\pm3.0)\times 10^7$ cm$^2$, where formal
$1\sigma$ statistical uncertainties in spectral fitting (i.e.,
neglecting possible systematic effects due to calibration
uncertainties) and distance have been propagated, and isotropic
emission over $4\pi$ sr has been assumed. The new parallactic distance
does not, by itself, alleviate the apparent discrepancy between
observed and expected polar cap sizes, increasing the inferred
emitting area by a factor $(361/250)^2 \simeq 2$.
 
Alternatively, the thermal fits of \citet{2003astro.ph..2452W}, to
{\em ROSAT} and {\em ASCA} data simultaneously, imply emission
luminosities of $L_1 = (1.85 \pm 0.30) \times 10^{30}$ ergs~\persec\
and $L_2 = (1.72 \pm 0.30)\times 10^{29}$ ergs~\persec, with areas
$A_1 = (1.88 \pm 0.35) \times 10^9$ cm$^2$ and $A_2 = (1.34 \pm 0.36)
\times 10^6$ cm$^2$, for the 2 MK and 7 MK blackbody components,
respectively. If this spectral model holds, the intrinsic emitting
area of the low-temperature component,
\begin{equation}
A^{\rm int}_1 = A_1 (1 - 2GM/Rc^2) = 1.1\times 10^9\,\,{\rm cm}^2,
\end{equation}
comes within a factor of three of the canonical polar-cap size defined
by the last open field lines in the pulsar magnetosphere,
\begin{equation}
A_{\rm pc} = \frac{2\pi^2 R^3}{cP} = 2.9\times 10^9\,\,{\rm cm}^2,
\end{equation}
where $P$ is the pulse period, an intriguing similarity given
uncertainties in magnetospheric geometry, the beaming solid angle,
and likely distortion of the surface blackbody spectrum due to
propagation of the radiation through a stellar atmosphere. 

If, instead, the observed X-ray emission is interpreted correctly by
the nonthermal model of \citet{2003astro.ph..2452W}, then the
0.1--10~keV luminosity implied by our new distance measurement is
$L_{\rm PL} = (9.6\pm0.8)\times 10^{30}$ ergs~\persec; in the 2--10
keV band, the same model implies a luminosity $3.1\times 10^{-4}$ of
the pulsar's spin-down power, $\dot E$, comparable to the nonthermal
X-ray emissions of other pulsars \citep{PCCM02}.

Finally, we address the implications of our new distance determination
on stellar cooling models for B1929+10. Following the procedure
adopted by \citet{YHH94}, we derive upper bounds on the surface
temperature by requiring that whole-surface ``cooling'' emission not
exceed the observed soft-band (0.09--0.30 keV) ROSAT countrate. The
result is sensitive to the assumed absorption column, $n_H$. For the
parallactic distance and the smallest likely $n_H$, $1.5\times
10^{20}$ cm$^{-2}$, a temperature $T \lesssim 0.24$ MK is required,
corresponding to a luminosity limit $L \lesssim 4\times 10^{30}$
ergs~s$^{-1}$, roughly an order of magnitude larger than typical
theoretical expectations ($\sim 5\times 10^{29}$ ergs~s$^{-1}$) for
residual heat retained since the formation of such an old pulsar
\citep{TTT+02}. If $n_H$ is allowed to be as large as $6.5\times
10^{20}$ cm$^{-2}$, the 1$\sigma$ upper limit for thermal models
derived by \citet{2003astro.ph..2452W}, an even hotter surface is
allowed: $T \lesssim 0.35$ MK and $L \lesssim 2\times 10^{31}$
ergs~s$^{-1}$. We therefore conclude that available soft X-ray
measurements do not usefully constrain leading cooling models.

\subsection{High Energy Emission from B0355+54}

PSR~B0355+54 was detected by the {\em ROSAT} observatory
\citep{S94}. Too few photons were acquired to permit detailed spectral
modeling, but the gross distribution of photon energies suggested
nonthermal emission. A recent observation with the {\em XMM-Newton}
telescope has confirmed this conclusion (Kennea et al., in
preparation). Adopting a distance of 2.1 kpc, \citet{S94} inferred a
nonthermal luminosity (0.1--2.4 keV) of $1.0\times 10^{32}$
ergs~\persec; scaling to the parallactic distance of 1.04 kpc results
in a revised luminosity of $2.4\times 10^{31}$ ergs~\persec. The
available {\em XMM-Newton} data also provide a $1\sigma$ upper bound
on the surface cooling luminosity of the neutron star, $L < 4.5\times
10^{32}$ ergs~\persec\ (J. Kennea, private communication).  As with
B1929+10, this limit is consistent with leading cooling models
\citep{TTT+02}.

The new distance measurement for B0355+54 has similar implications for
studies of the $\gamma$-ray emission from the pulsar.  The EGRET
instrument aboard {\em CGRO} set an upper limit to the flux of 0.1--10
GeV photons that implies a conversion efficiency $\eta_\gamma \equiv
L_\gamma/\dot E < 0.2$ \citep{NAB+96}, if the $\gamma$-ray beam
crosses the line of sight.  Our parallactic distance, nearly a factor
of two smaller than the dispersion-derived distance, further reduces
the apparent $\gamma$-ray efficiency, $\eta_\gamma < 0.05$, placing
B0355+54 at odds with the observed trend of increasing $\eta_\gamma$
with spindown age.  For example, PSR B1055$-$52, with a characteristic
age similar to B0355+54, has $\eta_\gamma \simeq 0.7$. If the trend of
$\gamma$-ray efficiency increasing with age holds, the likely
conclusion is that the $\gamma$-ray beam for B0355+54 does not cross
our line of sight, as predicted in the outer-gap beaming model of
\citet{YR95}.

\section{Conclusions}\label{end}

We have presented results from VLBA astrometric observations of three
pulsars at 5~GHz.  For B0329+54, the systematic errors are large
enough that we obtain only an upper limit to the parallax, $\pi <
1.5$~mas, and proper motions consistent with the \citet{BBGT02} value.
The large angular separation between the calibrator and pulsar
(5.3\arcdeg) is the most probable cause for the large systematic
errors. With the same calibrator and at the same observation epochs,
but with a smaller angular throw (3.3\arcdeg), we obtain a proper
motion and parallax for B0355+54, with an implied $D =
1.04^{+0.21}_{-0.16}$~kpc and $\vperp = 61^{+12}_{-9}$~km~\persec.

For B1929+10, we demonstrate the consistency of our absolute
astrometry with that of \citet{BBGT02} to within $\sim 2.5$~mas. The
longer time baseline of the 5~GHz data permits a better proper motion
measurement while reducing the covariance between $\mu$ and $\pi$.  A
joint fit to both data sets produces a parallax identical to that
measured with the 5~GHz data alone (but with smaller uncertainties),
implying a distance $D = 0.361^{+0.010}_{-0.008}$~kpc and velocity
$\vperp = 177^{+4}_{-5}$~km~\persec.  The results for B1929+10
illustrate the need for both a time baseline long enough to
minimize the covariance between $\mu, \pi$, and a sufficient number of 
epochs so that the parallax signature is well sampled.

Based on the observed astrometric errors in our observations, as well
as the astrometric uncertainties measured by \citet{CCL+01} and
\citet{VVD+03} at 1.4~GHz, we quantify the astrometric error as a
function of calibrator throw and observing frequency. The astrometric
precision is correlated with the angular throw between the calibrator
and the target, with lower errors at 5~GHz compared to
1.4~GHz. \Fref{errors} provides a first attempt at quantifying this
dependence on an empirical basis, although we note that the random
nature of these errors implies that a value of parallax smaller than
the predicted astrometric error can be still be measured given enough
measurement epochs.  The calibrator throw in each dimension may serve
as an important predictor of the accuracy attainable in that
dimension.  Future observations will allow us to confirm or refute
this apparent correlation.

Finally, we have investigated the implications of the new distance and
transverse velocity determinations for B1929+10 and B0355+54. We
conclude that B1929+10 and the runaway O-star $\zeta$~Ophiuchi do not
approach within 10~pc of each other when their paths are traced back
in time, given the revised distance and velocity for the pulsar. Thus,
contrary to the suggestion by \citet{HBZ01}, they are unlikely to have
been members of a binary system that was disrupted by the birth
supernova of the pulsar.  A review of X-ray data allows us to
constrain the luminosity of the two pulsars in soft X-rays. The limits
on the thermal emission from both objects are consistent with (and
unconstraining for) leading neutron star cooling models
\citep{TTT+02}.  If the X-ray emission from B1929+10 is thermal in
nature, the emitting area for X-rays is roughly consistent with the
canonical size for a heated polar cap.  For B0355+54, the inferred
conversion of spin-down power to gamma-ray luminosity is low, which
may imply that its $\gamma$-ray beam does not intersect our line of
sight.

The implications that we discuss here illustrate the range and
diversity of the applications of new distance and velocity
measurements for neutron stars.  Most pulsars are weak at 5~GHz, but
we expect that a few others will be observable with the VLBA
alone. The addition of telescopes with large collecting areas such as
Arecibo and the Green Bank Telescope to the array will allow us to
further enlarge the sample, and explore the consequences at multiple
wavelengths.

\acknowledgements

We thank Andrew Lyne and Michael Kramer at the Jodrell Bank
Observatory for pulsar timing solutions used to gate the VLBA
correlator, Walter Brisken for useful discussions about the 1.4~GHz
astrometric results, and Jamie Kennea for sharing results from XMM
observations of B0355+54 prior to publication.  The National Radio
Astronomy Observatory is a facility of the National Science Foundation
(NSF) operated under cooperative agreement by Associated Universities,
Inc.  This work at Cornell was supported in part by NSF grants AST
9819931 and AST 0206036, and made use of NASA's Astrophysics Data
System Abstract Service and the {\tt arXiv.org} astro-ph preprint
service, as well as data obtained from the High Energy Astrophysics
Science Archive Research Center (HEASARC), provided by NASA's Goddard
Space Flight Center.  WV thanks the Niels Stensen Foundation for
partially supporting his stay at Cornell University, and ZA
acknowledges support from grant NRA-99-01-LTSA-070 to NASA GSFC.
Basic research in radio astronomy at the NRL is supported by the
Office of Naval Research.


\end{document}